\documentclass[aps,prl,preprint,superscriptaddress,amsmath,amssymb]{revtex4-1}

\usepackage{graphicx,color}
\usepackage{float}
\usepackage{braket}
\usepackage[textsize=tiny]{todonotes}
\usepackage{changes}

\definecolor{vert}{rgb}{0.04,0.67,0.07}
\definecolor{mauve}{rgb}{0.22,0.33,0.64}

\begin{document}
\title{Taking advantage of light- and heavy-hole trions for optical spin initialization, control and readout}
\author{G. \'Ethier-Majcher}
\affiliation{D\'{e}partement de g\'{e}nie physique, Polytechnique Montr\'{e}al, Montr\'{e}al, H3C 3A7, Canada}
\author{P. St-Jean}
\affiliation{D\'{e}partement de g\'{e}nie physique, Polytechnique Montr\'{e}al, Montr\'{e}al, H3C 3A7, Canada}
\author{S. Francoeur}
\affiliation{D\'{e}partement de g\'{e}nie physique, Polytechnique Montr\'{e}al, Montr\'{e}al, H3C 3A7, Canada}

\begin{abstract}
Optical control strategies in semiconductor nanostructures have almost exclusively relied on heavy-hole exciton and trion states. In the first part of this letter, we show that light-hole trions provide the missing ressource for consolidating all single qubit operations in a mutually compatible magnetic field configuration: electron spin initialization and control can be achieved through light-hole trion states and cycling transition is provided by heavy-hole trion states. In the second part, we experimentally demonstrate that pairs of nitrogen atoms in GaAs exhibiting a $C_s$ symmetry bind both light- and heavy-hole excitons and negative trions. A detailed analysis of the fine structure reveals that that trion states provide the lambda level structure necessary for fast initialization and control along with energetically-protected cycling transition compatible with single-shot readout. 
 \end{abstract}

\maketitle
Optically-controlled qubits in semiconductor quantum dots have long been considered strong candidates for quantum information processing \cite{Imamoglu1999} and critical operations have been demonstrated using heavy-hole exciton and trion states: fast initialization \cite{Atature2006,Xu2007,Press2008}, high-fidelity control \cite{Press2008,DeGreve2011}, and robust single-shot readout \cite{Delteil2014}. However, a significant hurdle remains as these three operations require mutually exclusive magnetic field configurations \cite{Warburton2013,DeGreve2013}, preventing their sequential application and impeding further developments of the field.

Pushed to relatively high energy by strain and confinement, states derived from light-hole valence bands have often been discarded or considered undesirable: their proximity doubles the number of states involved in the analysis, alters heavy-hole selection rules \cite{Atature2006,Delteil2014,Tonin2012,Luo2014}, and opens additional relaxation channels \cite{Tsitsishvili2015}. Nonetheless, the angular composition of light-hole states and their associated selection rules could provide distinctive advantages. For example, direct quantum information transfers between photons and electrons \cite{Vrijen2001}, fast RF control of hole qubits \cite{Sleiter2006} and arbitrary electron qubit rotations through virtual excitations \cite{Dubin2008} are made possible and strain-engineered quantum dots have recently allowed making light-hole excitons the lowest energy states \cite{Huo2013}. Here, we demonstrate that simultaneous optical access to light- and heavy-hole states allows implementing all necessary qubit operations in a single magnetic field configuration, demonstrating that light holes provide an exceptional resource for quantum information processing in semiconductor nanostructures. Then, we show that nitrogen pairs in GaAs conveniently provide the optical access to light- and heavy-hole trions and the required energetic level structure to the realization of the proposed scheme without the need for complex strain-control nanofabrication steps. 
  
\begin{figure}[hbpt]
	\centering
	\includegraphics[width=9cm]{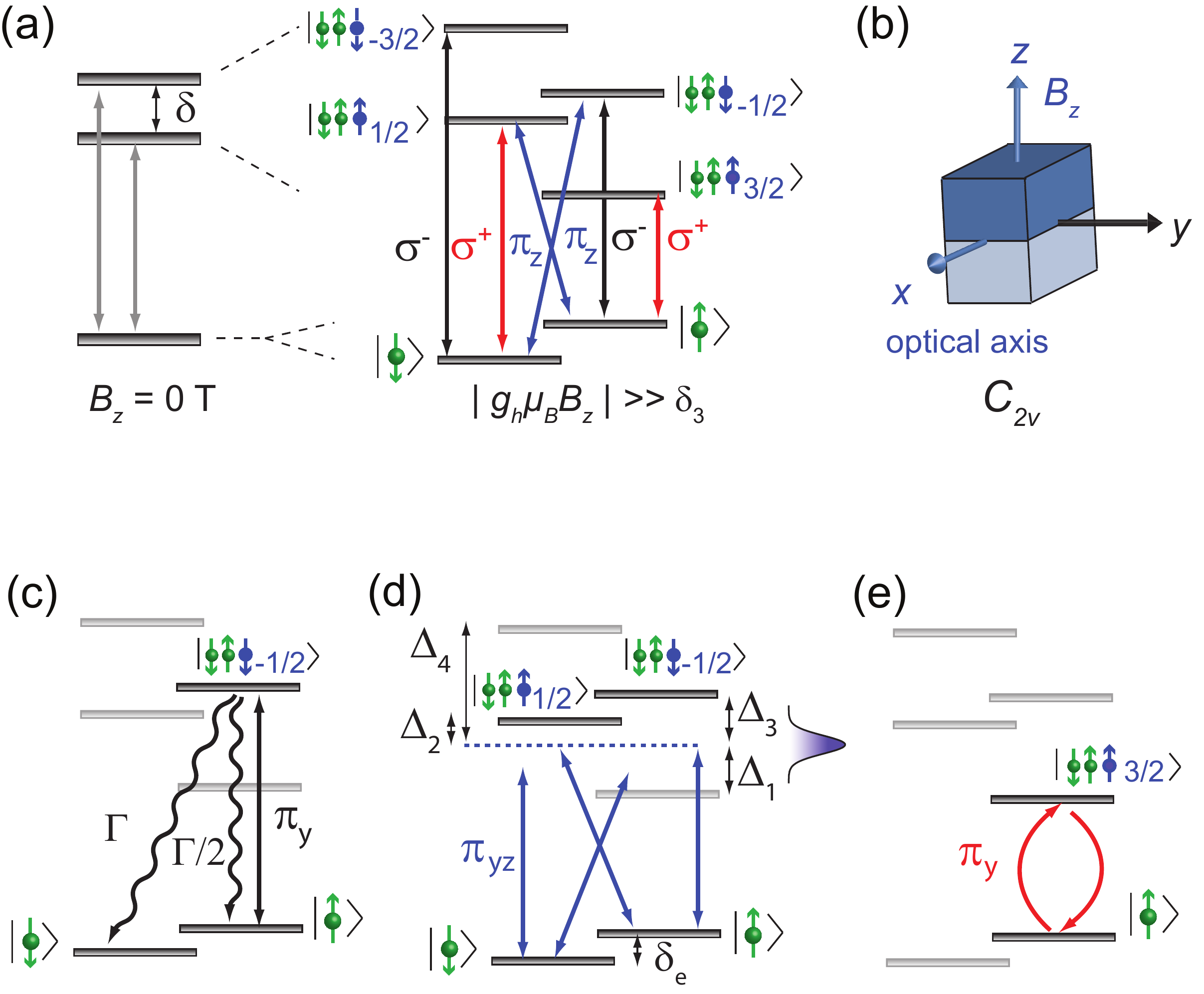}
	\caption{(color online) (a) Energy-level diagram at zero magnetic-field, showing the splitting of trion states $\delta$ due to the crystal field, and at a magnetic field for which the Zeeman interaction is dominant. Since electrons are in a spin singlet configuration, trion states are simply tagged by their hole angular momentum. Polarization selection rules are given : red and black arrows indicate circular polarization in the $xy$ plane and blue arrows linear polarizations along $z$. (b) Schematic representation of a nanostructure of $C_{2v}$ symmetry, with $z$ chosen as to minimize $|v_x -v_y|$ (which is not necessarily the direction of the $C_2$ axis). The optical axis is perpendicular to the magnetic field $B_z$. (c) Spin initialization: a linearly polarized laser resonantly pumps a light-hole trion transition to initialize the spin in the $|\color{vert}{\downarrow} \color{black}\rangle$ state. (d) Spin control: a rotation laser broader than the electron spin splitting $\delta_e$ drives the light-hole trion states with detuning $\Delta_i$. This results in stimulated Raman transitions. (e) Readout: a resonant linearly polarized laser drives a cycling heavy-hole transition.}
	\label{control}
\end{figure}

Our scheme takes advantage of the polarization selection rules connecting electron spin states to light- and heavy-hole trions. For this proposal, we consider a nanostructure of symmetry $C_{2v}$ described by three distinct crystal field coefficients $(v_x,v_y,v_z)$.  Figure \ref{control}(a) illustrates the electron and trion fine structures, where optical transitions occur between electron spin states $S_z=\pm1/2$ and four negative trion states $J_z=\pm1/2$ or $\pm3/2$. At zero field, the fine structure is dominated by the crystal field interaction, resulting in a splitting $\delta$ between trion states of mixed character, but nonetheless presenting either a dominant light- or heavy-hole component. Under a magnetic field applied along $z$, the Zeeman interaction competes with the crystal field, and trion states can be obtained by diagonalizing the following Hamiltonian expressed in the basis $J_z=\{3/2,1/2,-1/2,-3/2\}$, 

\small
\begin{align}
	H &= \sum_i{v_iJ_i^2} + \alpha_z J_z \\
	&=
	\begin{pmatrix}
		\delta_1+ \frac{3}{2}\alpha_z & 0 & \delta_3 & 0\\
		0 & \delta_2+\frac{1}{2}\alpha_z & 0 & \delta_3\\
		\delta_3 & 0 & \delta_2-\frac{1}{2}\alpha_z & 0\\
		0 & \delta_3 & 0 & \delta_1-\frac{3}{2}\alpha_z\\
	\end{pmatrix}\nonumber,
	\label{E-Hamiltonian}
\end{align}
\normalsize
\noindent where $\delta_1= \frac{3}{4}(3v_z +v_x+v_y)$, $\delta_2= \frac{1}{4}(v_z +7v_x+7v_y)$, $\delta_3= \frac{\sqrt{3}}{2}(v_x-v_y)$, $\delta=\sqrt{(\delta_1-\delta_2)^2+4\delta_3^2}$, and $\alpha_z=g_h \mu_B B_z$. The Zeeman Hamiltonian in this linear approximation depends only on the hole g-factor $g_h$ and the Bohr magneton $\mu_B$.

The key condition to the realization of this scheme is $\alpha_z\gg \delta_3$, such that the mixing of hole states due to off-diagonal terms is minimized. This is achieved by taking $z$ as the direction minimizing $|v_x-v_y|$ and by applying a sufficiently high field along this direction. When this condition is satisfied, trion states are eigenstates of the magnetic field and the fine structure shown in Fig. \ref{control}(a) is obtained. The polarization selection rules for the 6 allowed transitions are straightforward : heavy-hole transitions are circularly polarized ($\sigma^+$ and $\sigma^-$); light-hole transitions are either circularly polarized ($\sigma^+$ and $\sigma^-$) or linearly polarized ($\pi_z$). By taking the optical axis along $x$, as shown in Fig. \ref{control}(b), these selection rules can be used used to conveniently initialize, control and readout the spin.

The initialization and control procedures depicted in Fig.\ref{control}(c)-(d) take advantage of the double lambda system provided by light-hole trion states. Initialization is done by resonantly exciting the transition from $\ket{\color{vert}\uparrow\color{black}}$ to $\ket{\color{vert}\downarrow\uparrow\color{mauve}\Downarrow_{-1/2}\color{black}}$ using $\pi_y$ polarization. This trion state either decays to $\ket{\color{vert}\uparrow\color{black}}$ with rate $\Gamma/2$ or to $\ket{\color{vert}\downarrow\color{black}}$ with rate $\Gamma$. After a few cycles, the electron is initialized in $\ket{\color{vert}\downarrow\color{black}}$.
 
Fast and reliable coherent control can be realized using stimulated Raman transitions \cite{Press2008}, as depicted schematically in Fig. \ref{control}(d). A rotating laser broader than the electron spin splitting $\delta_e$ and detuned from all trion states ($\Delta_i\gg\Omega_i$, where $\Omega_i$ are the Rabi frequencies of the light-hole transitions) adiabatically rotates the electron state from $\ket{\color{vert}\downarrow\color{black}}$ to $\ket{\color{vert}\uparrow\color{black}}$. As shown in Supplemental Material, heavy-hole trions induce AC Stark shifts of electron levels but do not contribute to rotation. The effective Rabi frequency is maximum for a $\pi_{yz}$ rotating laser, i.e., linearly polarized in the $yz$ plane at $45^\circ$ with respect to the magnetic field. 

Free from an ancillary readout interface \cite{Vamivakas2010}, the spin readout scheme  shown in Fig.\ref{control}(e) is similar to the one applied in Ref. \onlinecite{Delteil2014}. A resonant $\pi_y$ laser excites one of the two heavy-hole trions. As any of these states only couples to a single electron state, a energetically-protected cycling transition is available for single-shot readout. Here, detection can be performed along the optical axis defined in Fig.\ref{control}(b) or alternatively along $z$ for better signal discrimination and collection efficiency. The maximum readout fidelity is evaluated from $1-\eta^{-1}$, where $\eta=\frac{3}{2}\left(\frac{2\alpha_z+\delta_1-\delta_2}{\delta_3}\right)^2$ is the branching ratio, defined as the square of the oscillator strength ratio between the cycling and forbidden transitions originating from a given heavy-hole trion state. 

Central to this proposal are three key conditions: 1) optical access to both light-and heavy-hole trions, 2) a Voigt configuration and 3) a relatively small $\delta_3$. If conditions 2 and 3 are rather straighforward to implement, condition 1 represents a significant obstacle. Strain-engineered quantum dots (QDs) \cite{Huo2013} are interesting candidates, but light-hole trions have yet to be observed. Recently, isoelectronic centers were proposed as promising alternatives to QDs for quantum information processing \cite{Ethier-Majcher2014} and, in the second part of this letter, we demonstrate that centers formed from a pair of nitrogen impurities in GaAs provide convenient optical access to both light- and heavy-hole trions. Furthermore, a detailed analysis of their emission reveals an energetic structure closely approaching the one shown in Fig. \ref{control}(a) for a magnetic field configuration minimizing mixing due to $\delta_3$. 

\begin{figure}[hbpt]
	\centering
	\includegraphics[width=9cm]{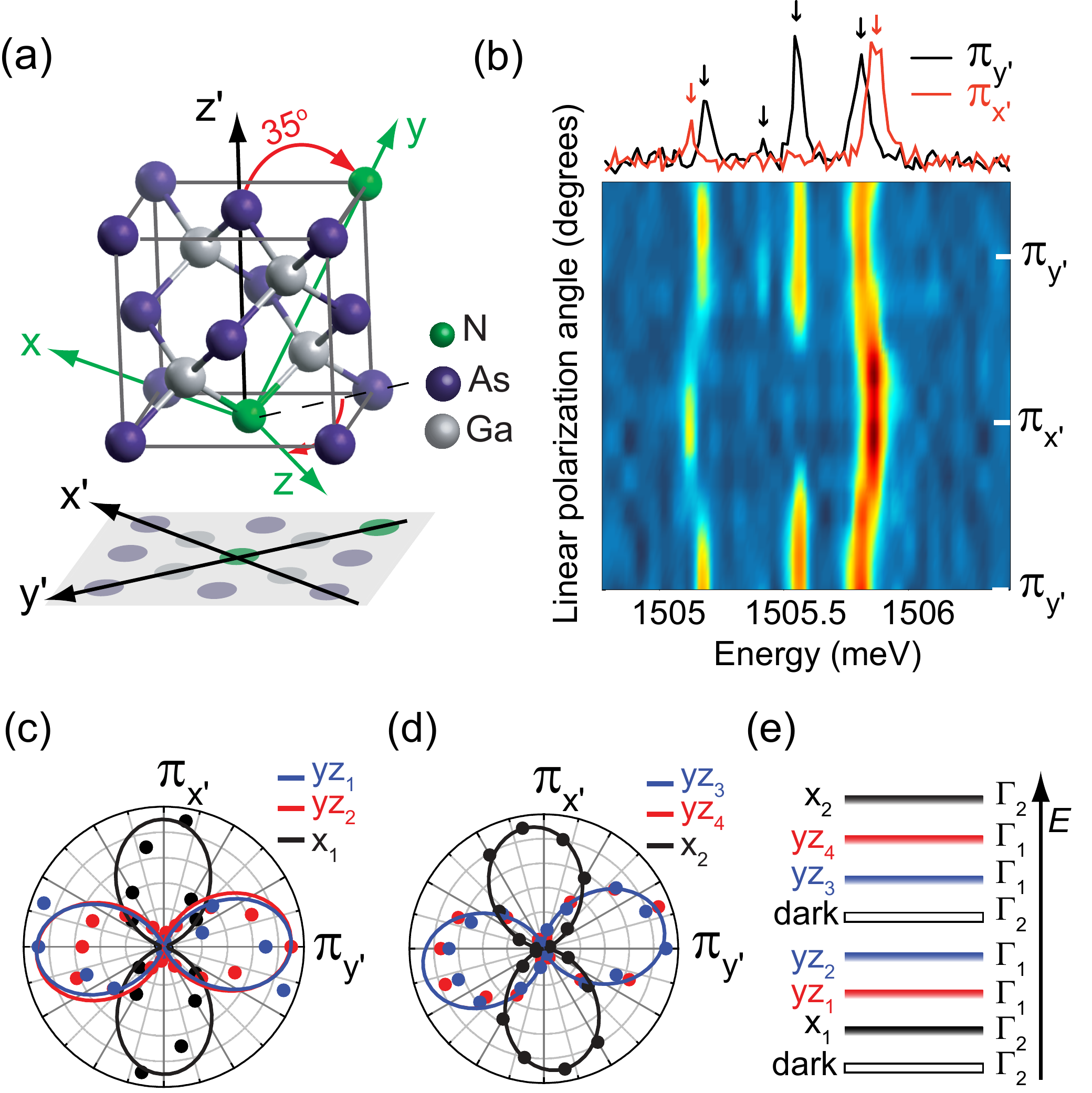}
	\caption{(color online) (a) Schematic representation of a nitrogen pair of $C_s$ symmetry oriented along [112], where $(x,y,z)$ are nitrogen pair coordinates and $(x',y',z')$ define the laboratory frame. The pair direction ($y$) is tilted by $35^\circ$ with respect to the optical axis ($z'$). (b) PL intensity as a function of energy and linear polarization angle for a bound exciton. Four transitions are polarized along $y'$ and two along $x'$. (c) Polar diagram showing the normalized PL intensities for the three low-energy transitions and (d) high-energy transitions. $yz_i$ denotes exciton states emitting linearly polarized light in the $yz$ plane and detected along $x'$, while $x_i$ denotes exciton states emitting linearly polarized light along $x$ and detected along $y'$ in the PL frame. (e) Schematic representations of the energy levels and their representation  for an exciton bound to a pair of $C_s$ symmetry (See Supplemental Material).}
	\label{exciton}
\end{figure}
 
 The studied sample was grown by molecular beam epitaxy along [001] and is composed of a 25~nm layer of GaAs:N clad by two 5~nm GaAs layers and sandwiched between two Al$_{0.25}$Ga$_{0.75}$As barriers. Microluminescence measurements were carried out at 4~K in a custom made confocal microscope. The sample was excited at 780~nm and the polarization was analyzed using a quarter or a half-wave-plate and a linear polarizer.  Magnetic fields from -6 to 6~T were applied parallel to the optical axis of the microscope and along the growth direction.

\begin{figure*}[hbpt]
	\centering
	\includegraphics[width=18cm]{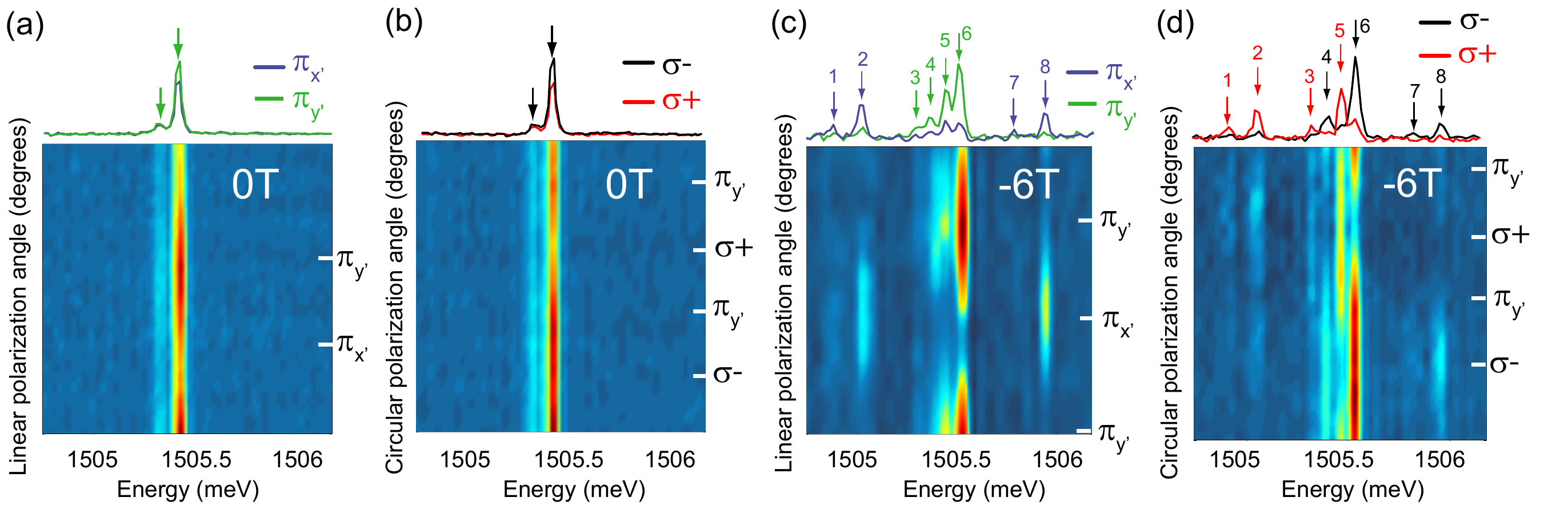}
	\caption{(color online) PL Intensity from a negative trion as a function of energy at (a) and (b) 0~T and (c) and (d) -6~T. PL is detected in a circular polarization basis in (a) and (c) and in a linear polarization basis in (b) and (d). Spectra are presented on top of each map for the given polarizations.}
	\label{pol} 
\end{figure*}

In contrast with other extensively studied $C_{2v}$ configurations in GaAs \cite{Francoeur2004, Marcet2009, Marcet2010, St-Jean2015}, the nitrogen pair configuration of $C_s$ symmetry binding negative trions has never been reported. Following the discussion presented in Supplemental Material, it is assigned to a defect involving two nitrogen atoms positioned along the [112] direction. This configuration is schematically represented in Fig. \ref{exciton}(a). Two sets of coordinates are used in the analysis of the data. The first set $(x,y,z)$ is defined with respect to the nitrogen pair: $x$ lies along the $[\bar{1}10]$ direction which is perpendicular to the mirror plane, $y$ is taken along the pair axis and $z$ along $[11\bar{1}]$. The measurement frame is defined through $(x',y',z')$, where $z'$ is along the optical axis of the photoluminescence (PL) microscope. $y'$ is along $[\bar{1}\bar{1}0]$ and corresponds to the projection of both $y$ and $z$ in the sample plane. Similarly, $x'$ is along $[\bar{1}10]$ and corresponds to the projection of $x$ in the sample plane.  Figure \ref{exciton}(b) shows the PL intensity from an exciton bound to this nitrogen pair as a function of energy and linear polarization angle of the emission. The fine structure results from the exchange interaction, the crystal field, and confinement effects and is composed of six linearly polarized lines. Figure \ref{exciton}(c) and d present respectively the normalized intensities of the three lower and higher energy transitions on a polar diagram. As can be seen, two transitions are linearly polarized along $x'$ and four along $y'$. The number of observed transitions and the polarization angles are all consistent with fine structure expected from a defect of $C_s$ symmetry. Although considerable mixing occurs between these states all located within an 1~meV window, it is still comparable to mixing observed in QDs \cite{Tonin2012,Luo2014} and the four low-energy (high-energy) states exhibit a dominant heavy-hole (light-hole) character (See Supplemental Material).

This $C_s$ nitrogen pair configuration can also bind trions. From the number of transitions observed, the diamagnetic shift and the pseudo-acceptor nature of the nitrogen pair, we conclude that the trions are negatively charged (see Supplemental Material for a detailed discussion). Figure \ref{pol} shows the emission fine structure of a negative trion at zero and high magnetic field, detected in linear (Fig. \ref{pol}(a)) and circular polarization basis (Fig. \ref{pol}(b)). As expected from fig. \ref{control}a, the fine structure is composed of two transitions at zero field split by $\delta=100$ $\mu$eV; the high (low) energy lines correspond to dominant light- (heavy-) hole trion states. The absence of clear polarization contrast in both the linear and circular bases indicates that these two lines are formed from several transitions. Indeed, they evolve into two quadruplets when a magnetic field is applied along $z'$ (Fig. \ref{pol}(c) and (d)). Each transition shows a preferential linear polarization either along $\pi_{x'}$ or $\pi_{y'}$. Moreover, four transition exhibit preferential right circular polarization $\sigma^+$ and four have a stronger left circular polarization $\sigma^-$. These polarizations result from the interplay between the crystal field of the nitrogen pair and the magnetic field. Because the field is not applied along one of the pair axes ($x,y$ or $z$), its effect on the trion states is not trivial and the level structure cannot be directly compared to Fig. \ref{control}(a).

Excitons and trions were never observed simultaneously on a given nitrogen pair, suggesting that residual electrons are bound to the pairs for a extended time as expected from the charge binding mechanism proposed in Ref. \onlinecite{Hopfield1966}. This residual electron can be provided by dopants in the vicinity of the nitrogen pair or in the surrounding AlGaAs barrier. Although the binding energy of trions could not be determined directly from a single pair, measurements on tens of exciton and trions reveal that the trions emission energy is $\sim 500$~$\mu$eV lower then that of excitons. 

\begin{figure*}[ht]
	\centering
	\includegraphics[width=18cm]{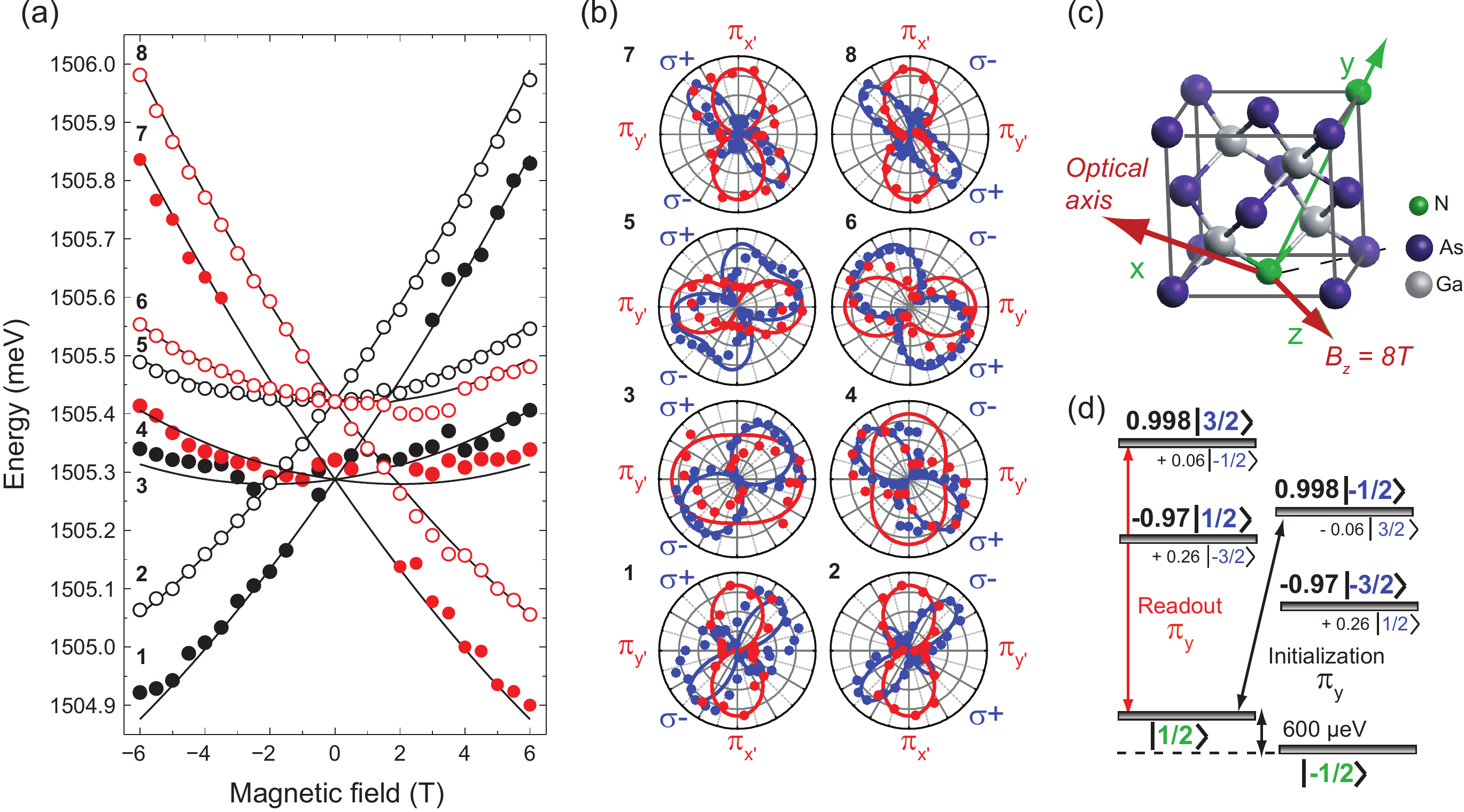}
	\caption{(color online) (a) Emission energies from a negative trion as a function of the magnetic field. Circles are the measured energies and solid lines are calculated. Solid (empty) circles represent states with a dominant heavy-hole (light-hole) component and transitions represented with red (black) circles are preferentially right (left) circularly polarized. (b) Normalized intensities in a polar diagram for every transition detected in circular (blue) and linear (red) polarization. (c) Optimal configuration for initialization, control and readout. The magnetic field is applied along $z$ and the optical axis is along $x$. (d) Calculated eigenstates for the magnetic field configuration of (c). Trion states (in blue) are only specified by their hole angular momentum to simplify the notation. Every state is dominated by an eigenstate of $J_z$ (coefficient in bold) and is weakly mixed. Electron eigenstates are given in green and are split by 600~µeV. The initialization (readout) transition is represented by a black (red) arrow.}
	\label{energies}
\end{figure*}

To evaluate the potential of this light- and heavy-hole trion for qubit operations presented in Fig. \ref{control}(c)-(e), a comprehensive analysis of the trion emission as a function of $B_{z'}$ and polarization is presented next. The initial state of the transition is modeled with the hole Hamiltonian, $H_i=H_{CF}+H_{Zh}$, where $H_{CF}$ describes the crystal field created by the nitrogen pair and $H_{Zh}$ the Zeeman effect. Using an invariant expansion, we can build an Hamiltonian for a system of $C_{s}$ symmetry \cite{Francoeur2010} with the crystal field expressed in terms of the coefficient $v_x$, $v_y$, $v_z$ and $v_{yz}$as follows, 
\begin{equation}
H_{CF}=\sum_{i=x,y,z} v_iJ_{i}^2+v_{yz}\{J_y,J_z\},
\end{equation}

\noindent where $x, y$ and $z$ are the nitrogen pair axes shown in fig. \ref{exciton}a and $J_i$ are the spin-$3/2$ Pauli matrices. The Zeeman Hamiltonian is 
\begin{equation}
	H_{Zh}=\mu_B\sum_{i=x,y,z}(K J_i+L J_i^3)B_i+CB^2,
\end{equation}

\noindent where $K$ and $L$ are coefficients for the isotropic and anisotropic Zeeman terms. To reduce the number of parameters involved, we assume that $K$, $L$ and the diamagnetic shift coefficient $C$ do not depend on the orientation of the magnetic field. The Zeeman Hamiltonian describing the final electron state is,
\begin{equation}
H_f=H_{Ze}=\frac{\mu_B}{2}\sum_{i=x,y,z}g_i\sigma_iB_i,
\end{equation}
where $g_i$ are the electron g-factors, which are considered anisotropic.

This Hamiltonian is used to reproduce the emission energies at all fields and the polarizations at -6 T for all 8 transitions. Figure \ref{energies}(a) shows a very good agreement between the calculated and measured energies. The red (black) circles represent transitions with dominant left (right) circular polarization and empty (filled) circles represent trion states with a higher (lower) light-hole proportion. Figure \ref{energies}(b) shows in a polar diagram the normalized intensity for transitions 1 to 8 (from low to high energy at -6~T) as a function of the linear polarization angle (red) and circular polarization angle (blue). Again, calculated polarizations (solid lines) are in good agreement with the experimental data (dots), except maybe for transitions 3 and 4, which are sensitive to slight tilts of the magnetic field with respect to $z'$, resulting in a linear polarizations aligned away from $x'$ and $y'$. The calculated energies and polarizations presented in Fig. \ref{energies} were obtained with $(v_x,v_y,v_z,v_{yz})=(10,-13,63,-1)$ $\mu$eV. The magnetic field parameters are $g_y=1.24$, $g_z=1.28$, $K=-1.16$, $L=0.78$ and $C=2.4~\mu \mathrm{eV/T}^2$. The electron g-factors are higher than the ones extracted for excitons bound to the pair emitting at 1508~meV \cite{Marcet2009}, indicating that an electron is more strongly bound when it is not forming an exciton. Because of the large value for $L$ and $K$, the effective g-factor is positive for heavy-holes and negative for light-holes. A significant anisotropic Zeeman interaction for holes was also reported for excitons bound to nitrogen pairs in GaAs and GaP, although weaker than the one reported here. The diamagnetic shift extracted from the data is similar to the ones measured for excitons bound to nitrogen pairs of $C_{2v}$ symmetry (1.99~$\mu$eV) and confirms that the extra charge is negative. 

This physical model is now used to evaluate the merit of this center of $C_s$ symmetry for the proposed initialization, control and readout scheme. From the crystal field parameters, the magnetic field should be applied along $z$ to minimize $\delta_3$ and the optical axis is taken along $x$ (Fig. \ref{energies}(c)). Figure \ref{energies}(d) presents the resulting eigenstates. At $B_z=8$~T, these are approaching the eigenstates of $J_z$ given in Fig.\ref{control}(a). The radiative decay rate is expected to be close to the one extracted for the exciton ($\Gamma=15$~ns$^{-1}$ from Ref.\onlinecite{St-Jean2015}) and should lead to fast initialization. Although the initialization transition shown in Fig.\ref{energies}(d) is not polarization selective, the energy selection should allow a high-fidelity spin initialization given that the closest transition lies $\sim150$~$\mu$eV at higher energy \cite{Xu2007}. The splitting between electron states is $600$~$\mu$eV and the spin control should be realized with a $\sim1$~meV broad laser. The quality of the readout can be defined by the branching ratio $\eta$ between the allowed heavy-hole transition used for cycling and the corresponding forbidden transition. We find $\eta=400$ at 8~T, which is comparable to the oscillator strength ratio of 450 found from single-shot readouts realized in QDs \cite{Delteil2014}. Accounting for the range of extracted crystal field parameters, we estimate that $100<\eta<1000$ for the studied nitrogen pair. Given that this value increases with the magnetic field, concurrent spin control and single-shot readout is realizable. It is interesting to note that this ratio can be significantly enhanced be selecting centers of higher symmetry: for $D_{2d}$ and $C_{3v}$, $ \delta_3\rightarrow 0$ and $\eta$ becomes very large. Such centers are available in GaP:N and other isoelectronic impurity-host systems.

To conclude, simultaneously exploiting light- and heavy-hole selection rules enables single qubit operations in a single magnetic field configuration.  Isoelectronic centres, with their large dipole moments\cite{Ethier-Majcher2014} and the high optical homogeneity conferred by their atomic size, may prove to be convenient and powerful building blocks for implementing optically-controlled quantum information processing.


\begin{thebibliography}{23}%
	\makeatletter
	\providecommand \@ifxundefined [1]{%
		\@ifx{#1\undefined}
	}%
	\providecommand \@ifnum [1]{%
		\ifnum #1\expandafter \@firstoftwo
		\else \expandafter \@secondoftwo
		\fi
	}%
	\providecommand \@ifx [1]{%
		\ifx #1\expandafter \@firstoftwo
		\else \expandafter \@secondoftwo
		\fi
	}%
	\providecommand \natexlab [1]{#1}%
	\providecommand \enquote  [1]{``#1''}%
	\providecommand \bibnamefont  [1]{#1}%
	\providecommand \bibfnamefont [1]{#1}%
	\providecommand \citenamefont [1]{#1}%
	\providecommand \href@noop [0]{\@secondoftwo}%
	\providecommand \href [0]{\begingroup \@sanitize@url \@href}%
	\providecommand \@href[1]{\@@startlink{#1}\@@href}%
	\providecommand \@@href[1]{\endgroup#1\@@endlink}%
	\providecommand \@sanitize@url [0]{\catcode `\\12\catcode `\$12\catcode
		`\&12\catcode `\#12\catcode `\^12\catcode `\_12\catcode `\%12\relax}%
	\providecommand \@@startlink[1]{}%
	\providecommand \@@endlink[0]{}%
	\providecommand \url  [0]{\begingroup\@sanitize@url \@url }%
	\providecommand \@url [1]{\endgroup\@href {#1}{\urlprefix }}%
	\providecommand \urlprefix  [0]{URL }%
	\providecommand \Eprint [0]{\href }%
	\providecommand \doibase [0]{http://dx.doi.org/}%
	\providecommand \selectlanguage [0]{\@gobble}%
	\providecommand \bibinfo  [0]{\@secondoftwo}%
	\providecommand \bibfield  [0]{\@secondoftwo}%
	\providecommand \translation [1]{[#1]}%
	\providecommand \BibitemOpen [0]{}%
	\providecommand \bibitemStop [0]{}%
	\providecommand \bibitemNoStop [0]{.\EOS\space}%
	\providecommand \EOS [0]{\spacefactor3000\relax}%
	\providecommand \BibitemShut  [1]{\csname bibitem#1\endcsname}%
	\let\auto@bib@innerbib\@empty
	%</preamble>
	\bibitem [{\citenamefont {Imamoglu, A.}\
		\emph {et~al.}(1999)\citenamefont
		{Imamoglu, A.}, \citenamefont {Awschalom},
		\citenamefont {Burkard}, \citenamefont {DiVincenzo}, \citenamefont {Loss},
		\citenamefont {Sherwin},\ and\ \citenamefont {Small}}]{Imamoglu1999}%
	\BibitemOpen
	\bibfield  {author} {\bibinfo {author} {\bibfnamefont {A.}~\bibnamefont
			{Imamoglu}}, \bibinfo {author}
		{\bibfnamefont {D.~D.}\ \bibnamefont {Awschalom}}, \bibinfo {author}
		{\bibfnamefont {G.}~\bibnamefont {Burkard}}, \bibinfo {author} {\bibfnamefont
			{D.~P.}\ \bibnamefont {DiVincenzo}}, \bibinfo {author} {\bibfnamefont
			{D.}~\bibnamefont {Loss}}, \bibinfo {author} {\bibfnamefont {M.}~\bibnamefont
			{Sherwin}}, \ and\ \bibinfo {author} {\bibfnamefont {A.}~\bibnamefont
			{Small}},\ }\href {\doibase 10.1103/PhysRevLett.83.4204} {\bibfield
		{journal} {\bibinfo  {journal} {Phys. Rev. Lett.}\ }\textbf {\bibinfo
			{volume} {83}},\ \bibinfo {pages} {4204} (\bibinfo {year}
		{1999})}\BibitemShut {NoStop}%
	\bibitem [{\citenamefont {Atat\"{u}re}\ \emph {et~al.}(2006)\citenamefont
		{Atat\"{u}re}, \citenamefont {Dreiser}, \citenamefont {Badolato},
		\citenamefont {H\"{o}gele}, \citenamefont {Karrai},\ and\ \citenamefont
		{Imamoglu}}]{Atature2006}%
	\BibitemOpen
	\bibfield  {author} {\bibinfo {author} {\bibfnamefont {M.}~\bibnamefont
			{Atat\"{u}re}}, \bibinfo {author} {\bibfnamefont {J.}~\bibnamefont
			{Dreiser}}, \bibinfo {author} {\bibfnamefont {A.}~\bibnamefont {Badolato}},
		\bibinfo {author} {\bibfnamefont {A.}~\bibnamefont {H\"{o}gele}}, \bibinfo
		{author} {\bibfnamefont {K.}~\bibnamefont {Karrai}}, \ and\ \bibinfo {author}
		{\bibfnamefont {A.}~\bibnamefont {Imamoglu}},\ }\href {\doibase
		10.1126/science.1126074} {\bibfield  {journal} {\bibinfo  {journal} {Science
				(New York, N.Y.)}\ }\textbf {\bibinfo {volume} {312}},\ \bibinfo {pages}
		{551} (\bibinfo {year} {2006})}\BibitemShut {NoStop}%
	\bibitem [{\citenamefont {Xu}\ \emph {et~al.}(2007)\citenamefont {Xu},
		\citenamefont {Wu}, \citenamefont {Sun}, \citenamefont {Huang}, \citenamefont
		{Cheng}, \citenamefont {Steel}, \citenamefont {Bracker}, \citenamefont
		{Gammon}, \citenamefont {Emary},\ and\ \citenamefont {Sham}}]{Xu2007}%
	\BibitemOpen
	\bibfield  {author} {\bibinfo {author} {\bibfnamefont {X.}~\bibnamefont
			{Xu}}, \bibinfo {author} {\bibfnamefont {Y.}~\bibnamefont {Wu}}, \bibinfo
		{author} {\bibfnamefont {B.}~\bibnamefont {Sun}}, \bibinfo {author}
		{\bibfnamefont {Q.}~\bibnamefont {Huang}}, \bibinfo {author} {\bibfnamefont
			{J.}~\bibnamefont {Cheng}}, \bibinfo {author} {\bibfnamefont {D.~G.}\
			\bibnamefont {Steel}}, \bibinfo {author} {\bibfnamefont {A.~S.}\ \bibnamefont
			{Bracker}}, \bibinfo {author} {\bibfnamefont {D.}~\bibnamefont {Gammon}},
		\bibinfo {author} {\bibfnamefont {C.}~\bibnamefont {Emary}}, \ and\ \bibinfo
		{author} {\bibfnamefont {L.~J.}\ \bibnamefont {Sham}},\ }\href {\doibase
		10.1103/PhysRevLett.99.097401} {\bibfield  {journal} {\bibinfo  {journal}
			{Physical Review Letters}\ }\textbf {\bibinfo {volume} {99}},\ \bibinfo
		{pages} {097401} (\bibinfo {year} {2007})}\BibitemShut {NoStop}%
	\bibitem [{\citenamefont {Press}\ \emph {et~al.}(2008)\citenamefont {Press},
		\citenamefont {Ladd}, \citenamefont {Zhang},\ and\ \citenamefont
		{Yamamoto}}]{Press2008}%
	\BibitemOpen
	\bibfield  {author} {\bibinfo {author} {\bibfnamefont {D.}~\bibnamefont
			{Press}}, \bibinfo {author} {\bibfnamefont {T.~D.}\ \bibnamefont {Ladd}},
		\bibinfo {author} {\bibfnamefont {B.}~\bibnamefont {Zhang}}, \ and\ \bibinfo
		{author} {\bibfnamefont {Y.}~\bibnamefont {Yamamoto}},\ }\href {\doibase
		10.1038/nature07530} {\bibfield  {journal} {\bibinfo  {journal} {Nature}\
		}\textbf {\bibinfo {volume} {456}},\ \bibinfo {pages} {218} (\bibinfo {year}
		{2008})}\BibitemShut {NoStop}%
	\bibitem [{\citenamefont {{De Greve}}\ \emph {et~al.}(2011)\citenamefont {{De
				Greve}}, \citenamefont {McMahon}, \citenamefont {Press}, \citenamefont
		{Ladd}, \citenamefont {Bisping}, \citenamefont {Schneider}, \citenamefont
		{Kamp}, \citenamefont {Worschech}, \citenamefont {H\"{o}fling}, \citenamefont
		{Forchel},\ and\ \citenamefont {Yamamoto}}]{DeGreve2011}%
	\BibitemOpen
	\bibfield  {author} {\bibinfo {author} {\bibfnamefont {K.}~\bibnamefont {{De
					Greve}}}, \bibinfo {author} {\bibfnamefont {P.~L.}\ \bibnamefont {McMahon}},
		\bibinfo {author} {\bibfnamefont {D.}~\bibnamefont {Press}}, \bibinfo
		{author} {\bibfnamefont {T.~D.}\ \bibnamefont {Ladd}}, \bibinfo {author}
		{\bibfnamefont {D.}~\bibnamefont {Bisping}}, \bibinfo {author} {\bibfnamefont
			{C.}~\bibnamefont {Schneider}}, \bibinfo {author} {\bibfnamefont
			{M.}~\bibnamefont {Kamp}}, \bibinfo {author} {\bibfnamefont {L.}~\bibnamefont
			{Worschech}}, \bibinfo {author} {\bibfnamefont {S.}~\bibnamefont
			{H\"{o}fling}}, \bibinfo {author} {\bibfnamefont {A.}~\bibnamefont
			{Forchel}}, \ and\ \bibinfo {author} {\bibfnamefont {Y.}~\bibnamefont
			{Yamamoto}},\ }\href {\doibase 10.1038/nphys2078} {\bibfield  {journal}
		{\bibinfo  {journal} {Nature Physics}\ }\textbf {\bibinfo {volume} {7}},\
		\bibinfo {pages} {872} (\bibinfo {year} {2011})}\BibitemShut {NoStop}%
	\bibitem [{\citenamefont {Delteil}\ \emph {et~al.}(2014)\citenamefont
		{Delteil}, \citenamefont {Gao}, \citenamefont {Fallahi}, \citenamefont
		{Miguel-Sanchez},\ and\ \citenamefont {Imamoglu}}]{Delteil2014}%
	\BibitemOpen
	\bibfield  {author} {\bibinfo {author} {\bibfnamefont {A.}~\bibnamefont
			{Delteil}}, \bibinfo {author} {\bibfnamefont {W.-B.}\ \bibnamefont {Gao}},
		\bibinfo {author} {\bibfnamefont {P.}~\bibnamefont {Fallahi}}, \bibinfo
		{author} {\bibfnamefont {J.}~\bibnamefont {Miguel-Sanchez}}, \ and\ \bibinfo
		{author} {\bibfnamefont {A.}~\bibnamefont {Imamoglu}},\ }\href {\doibase
		10.1103/PhysRevLett.112.116802} {\bibfield  {journal} {\bibinfo  {journal}
			{Physical Review Letters}\ }\textbf {\bibinfo {volume} {112}},\ \bibinfo
		{pages} {116802} (\bibinfo {year} {2014})}\BibitemShut {NoStop}%
	\bibitem [{\citenamefont {Warburton}(2013)}]{Warburton2013}%
	\BibitemOpen
	\bibfield  {author} {\bibinfo {author} {\bibfnamefont {R.~J.}\ \bibnamefont
			{Warburton}},\ }\href {\doibase 10.1038/nmat3585} {\bibfield  {journal}
		{\bibinfo  {journal} {Nature materials}\ }\textbf {\bibinfo {volume} {12}},\
		\bibinfo {pages} {483} (\bibinfo {year} {2013})}\BibitemShut {NoStop}%
	\bibitem [{\citenamefont {{De Greve}}\ \emph {et~al.}(2013)\citenamefont {{De
				Greve}}, \citenamefont {Press}, \citenamefont {McMahon},\ and\ \citenamefont
		{Yamamoto}}]{DeGreve2013}%
	\BibitemOpen
	\bibfield  {author} {\bibinfo {author} {\bibfnamefont {K.}~\bibnamefont {{De
					Greve}}}, \bibinfo {author} {\bibfnamefont {D.}~\bibnamefont {Press}},
		\bibinfo {author} {\bibfnamefont {P.~L.}\ \bibnamefont {McMahon}}, \ and\
		\bibinfo {author} {\bibfnamefont {Y.}~\bibnamefont {Yamamoto}},\ }\href
	{\doibase 10.1088/0034-4885/76/9/092501} {\bibfield  {journal} {\bibinfo
			{journal} {Reports on progress in physics. Physical Society (Great Britain)}\
		}\textbf {\bibinfo {volume} {76}},\ \bibinfo {pages} {092501} (\bibinfo
		{year} {2013})}\BibitemShut {NoStop}%
	\bibitem [{\citenamefont {Tonin}\ \emph {et~al.}(2012)\citenamefont {Tonin},
		\citenamefont {Hostein}, \citenamefont {Voliotis}, \citenamefont {Grousson},
		\citenamefont {Lemaitre},\ and\ \citenamefont {Martinez}}]{Tonin2012}%
	\BibitemOpen
	\bibfield  {author} {\bibinfo {author} {\bibfnamefont {C.}~\bibnamefont
			{Tonin}}, \bibinfo {author} {\bibfnamefont {R.}~\bibnamefont {Hostein}},
		\bibinfo {author} {\bibfnamefont {V.}~\bibnamefont {Voliotis}}, \bibinfo
		{author} {\bibfnamefont {R.}~\bibnamefont {Grousson}}, \bibinfo {author}
		{\bibfnamefont {A.}~\bibnamefont {Lemaitre}}, \ and\ \bibinfo {author}
		{\bibfnamefont {A.}~\bibnamefont {Martinez}},\ }\href {\doibase
		10.1103/PhysRevB.85.155303} {\bibfield  {journal} {\bibinfo  {journal}
			{Physical Review B}\ }\textbf {\bibinfo {volume} {85}},\ \bibinfo {pages}
		{155303} (\bibinfo {year} {2012})}\BibitemShut {NoStop}%
	\bibitem [{\citenamefont {Luo}\ \emph {et~al.}(2014)\citenamefont {Luo},
		\citenamefont {Bester},\ and\ \citenamefont {Zunger}}]{Luo2014}%
	\BibitemOpen
	\bibfield  {author} {\bibinfo {author} {\bibfnamefont {J.-W.}\ \bibnamefont
			{Luo}}, \bibinfo {author} {\bibfnamefont {G.}~\bibnamefont {Bester}}, \ and\
		\bibinfo {author} {\bibfnamefont {A.}~\bibnamefont {Zunger}},\ }\href@noop {}
	{\bibfield  {journal} {\bibinfo  {journal} {arXiv:1411.6187}\ } (\bibinfo
		{year} {2014})}\BibitemShut {NoStop}%
	\bibitem [{\citenamefont {Tsitsishvili}(2015)}]{Tsitsishvili2015}%
	\BibitemOpen
	\bibfield  {author} {\bibinfo {author} {\bibfnamefont {E.}~\bibnamefont
			{Tsitsishvili}},\ }\href {\doibase 10.1103/PhysRevB.91.155434} {\bibfield
		{journal} {\bibinfo  {journal} {Physical Review B}\ }\textbf {\bibinfo
			{volume} {91}},\ \bibinfo {pages} {155434} (\bibinfo {year}
		{2015})}\BibitemShut {NoStop}%
	\bibitem [{\citenamefont {Vrijen}\ and\ \citenamefont
		{Yablonovitch}(2001)}]{Vrijen2001}%
	\BibitemOpen
	\bibfield  {author} {\bibinfo {author} {\bibfnamefont {R.}~\bibnamefont
			{Vrijen}}\ and\ \bibinfo {author} {\bibfnamefont {E.}~\bibnamefont
			{Yablonovitch}},\ }\href {\doibase 10.1016/S1386-9477(00)00296-4} {\bibfield
		{journal} {\bibinfo  {journal} {Physica E}\ }\textbf {\bibinfo {volume}
			{10}},\ \bibinfo {pages} {569} (\bibinfo {year} {2001})}\BibitemShut
	{NoStop}%
	\bibitem [{\citenamefont {Sleiter}\ and\ \citenamefont
		{Brinkman}(2006)}]{Sleiter2006}%
	\BibitemOpen
	\bibfield  {author} {\bibinfo {author} {\bibfnamefont {D.}~\bibnamefont
			{Sleiter}}\ and\ \bibinfo {author} {\bibfnamefont {W.~F.}\ \bibnamefont
			{Brinkman}},\ }\href {\doibase 10.1103/PhysRevB.74.153312} {\bibfield
		{journal} {\bibinfo  {journal} {Physical Review B}\ }\textbf {\bibinfo
			{volume} {74}},\ \bibinfo {pages} {153312} (\bibinfo {year}
		{2006})}\BibitemShut {NoStop}%
	\bibitem [{\citenamefont {Dubin}\ \emph {et~al.}(2008)\citenamefont {Dubin},
		\citenamefont {Combescot}, \citenamefont {Brennen},\ and\ \citenamefont
		{Melet}}]{Dubin2008}%
	\BibitemOpen
	\bibfield  {author} {\bibinfo {author} {\bibfnamefont {F.}~\bibnamefont
			{Dubin}}, \bibinfo {author} {\bibfnamefont {M.}~\bibnamefont {Combescot}},
		\bibinfo {author} {\bibfnamefont {G. K.}~\bibnamefont {Brennen}}, \ and\
		\bibinfo {author} {\bibfnamefont {R.}~\bibnamefont {Melet}},\ }\href
	{\doibase 10.1103/PhysRevLett.101.217403} {\bibfield  {journal} {\bibinfo
			{journal} {Physical Review Letters}\ }\textbf {\bibinfo {volume} {101}},\
		\bibinfo {pages} {217403} (\bibinfo {year} {2008})}\BibitemShut {NoStop}%
	\bibitem [{\citenamefont {Huo}\ \emph {et~al.}(2013)\citenamefont {Huo},
		\citenamefont {Witek}, \citenamefont {Kumar}, \citenamefont {Cardenas},
		\citenamefont {Zhang}, \citenamefont {Akopian}, \citenamefont {Singh},
		\citenamefont {Zallo}, \citenamefont {Grifone}, \citenamefont {Kriegner},
		\citenamefont {Trotta}, \citenamefont {Ding}, \citenamefont {Stangl},
		\citenamefont {Zwiller}, \citenamefont {Bester}, \citenamefont {Rastelli},\
		and\ \citenamefont {Schmidt}}]{Huo2013}%
	\BibitemOpen
	\bibfield  {author} {\bibinfo {author} {\bibfnamefont {Y.~H.}\ \bibnamefont
			{Huo}}, \bibinfo {author} {\bibfnamefont {B.~J.}\ \bibnamefont {Witek}},
		\bibinfo {author} {\bibfnamefont {S.}~\bibnamefont {Kumar}}, \bibinfo
		{author} {\bibfnamefont {J.~R.}\ \bibnamefont {Cardenas}}, \bibinfo {author}
		{\bibfnamefont {J.~X.}\ \bibnamefont {Zhang}}, \bibinfo {author}
		{\bibfnamefont {N.}~\bibnamefont {Akopian}}, \bibinfo {author} {\bibfnamefont
			{R.}~\bibnamefont {Singh}}, \bibinfo {author} {\bibfnamefont
			{E.}~\bibnamefont {Zallo}}, \bibinfo {author} {\bibfnamefont
			{R.}~\bibnamefont {Grifone}}, \bibinfo {author} {\bibfnamefont
			{D.}~\bibnamefont {Kriegner}}, \bibinfo {author} {\bibfnamefont
			{R.}~\bibnamefont {Trotta}}, \bibinfo {author} {\bibfnamefont
			{F.}~\bibnamefont {Ding}}, \bibinfo {author} {\bibfnamefont {J.}~\bibnamefont
			{Stangl}}, \bibinfo {author} {\bibfnamefont {V.}~\bibnamefont {Zwiller}},
		\bibinfo {author} {\bibfnamefont {G.}~\bibnamefont {Bester}}, \bibinfo
		{author} {\bibfnamefont {a.}~\bibnamefont {Rastelli}}, \ and\ \bibinfo
		{author} {\bibfnamefont {O.~G.}\ \bibnamefont {Schmidt}},\ }\href {\doibase
		10.1038/nphys2799} {\bibfield  {journal} {\bibinfo  {journal} {Nature
				Physics}\ }\textbf {\bibinfo {volume} {10}},\ \bibinfo {pages} {46} (\bibinfo
		{year} {2013})}\BibitemShut {NoStop}%
	\bibitem [{\citenamefont {Vamivakas}\ \emph {et~al.}(2010)\citenamefont
		{Vamivakas}, \citenamefont {Lu}, \citenamefont {Matthiesen}, \citenamefont
		{Zhao}, \citenamefont {F\"{a}lt}, \citenamefont {Badolato},\ and\
		\citenamefont {Atat\"{u}re}}]{Vamivakas2010}%
	\BibitemOpen
	\bibfield  {author} {\bibinfo {author} {\bibfnamefont {A.~N.}\ \bibnamefont
			{Vamivakas}}, \bibinfo {author} {\bibfnamefont {C.-Y. C.~C.}\ \bibnamefont
			{Lu}}, \bibinfo {author} {\bibfnamefont {C.}~\bibnamefont {Matthiesen}},
		\bibinfo {author} {\bibfnamefont {Y.}~\bibnamefont {Zhao}}, \bibinfo {author}
		{\bibfnamefont {S.}~\bibnamefont {F\"{a}lt}}, \bibinfo {author}
		{\bibfnamefont {A.}~\bibnamefont {Badolato}}, \ and\ \bibinfo {author}
		{\bibfnamefont {M.}~\bibnamefont {Atat\"{u}re}},\ }\href {\doibase
		10.1038/nature09359} {\bibfield  {journal} {\bibinfo  {journal} {Nature}\
		}\textbf {\bibinfo {volume} {467}},\ \bibinfo {pages} {297} (\bibinfo {year}
		{2010})}\BibitemShut {NoStop}%
	\bibitem [{\citenamefont {\'{E}thier-Majcher}\ \emph
		{et~al.}(2014)\citenamefont {\'{E}thier-Majcher}, \citenamefont {St-Jean},
		\citenamefont {Boso}, \citenamefont {Tosi}, \citenamefont {Klem},\ and\
		\citenamefont {Francoeur}}]{Ethier-Majcher2014}%
	\BibitemOpen
	\bibfield  {author} {\bibinfo {author} {\bibfnamefont {G.}~\bibnamefont
			{\'{E}thier-Majcher}}, \bibinfo {author} {\bibfnamefont {P.}~\bibnamefont
			{St-Jean}}, \bibinfo {author} {\bibfnamefont {G.}~\bibnamefont {Boso}},
		\bibinfo {author} {\bibfnamefont {A.}~\bibnamefont {Tosi}}, \bibinfo {author}
		{\bibfnamefont {J.~F.}\ \bibnamefont {Klem}}, \ and\ \bibinfo {author}
		{\bibfnamefont {S.}~\bibnamefont {Francoeur}},\ }\href {\doibase
		10.1038/ncomms4980} {\bibfield  {journal} {\bibinfo  {journal} {Nature
				Communications}\ }\textbf {\bibinfo {volume} {5}},\ \bibinfo {pages} {4980}
		(\bibinfo {year} {2014})}\BibitemShut {NoStop}%
	\bibitem [{\citenamefont {Francoeur}\ \emph {et~al.}(2004)\citenamefont
		{Francoeur}, \citenamefont {Klem},\ and\ \citenamefont
		{Mascarenhas}}]{Francoeur2004}%
	\BibitemOpen
	\bibfield  {author} {\bibinfo {author} {\bibfnamefont {S.}~\bibnamefont
			{Francoeur}}, \bibinfo {author} {\bibfnamefont {J.~F.}\ \bibnamefont {Klem}},
		\ and\ \bibinfo {author} {\bibfnamefont {A.}~\bibnamefont {Mascarenhas}},\
	}\href {\doibase 10.1103/PhysRevLett.93.067403} {\bibfield  {journal}
	{\bibinfo  {journal} {Physical Review Letters}\ }\textbf {\bibinfo {volume}
		{93}},\ \bibinfo {pages} {067403} (\bibinfo {year} {2004})}\BibitemShut
{NoStop}%
\bibitem [{\citenamefont {Marcet}\ \emph {et~al.}(2009)\citenamefont {Marcet},
	\citenamefont {Ouellet-Plamondon}, \citenamefont {Klem},\ and\ \citenamefont
	{Francoeur}}]{Marcet2009}%
\BibitemOpen
\bibfield  {author} {\bibinfo {author} {\bibfnamefont {S.}~\bibnamefont
		{Marcet}}, \bibinfo {author} {\bibfnamefont {C.}~\bibnamefont
		{Ouellet-Plamondon}}, \bibinfo {author} {\bibfnamefont {J.~F.}\ \bibnamefont
		{Klem}}, \ and\ \bibinfo {author} {\bibfnamefont {S.}~\bibnamefont
		{Francoeur}},\ }\href {\doibase 10.1103/PhysRevB.80.245404} {\bibfield
	{journal} {\bibinfo  {journal} {Physical Review B}\ }\textbf {\bibinfo
		{volume} {80}},\ \bibinfo {pages} {245404} (\bibinfo {year}
	{2009})}\BibitemShut {NoStop}%
\bibitem [{\citenamefont {Marcet}\ \emph {et~al.}(2010)\citenamefont {Marcet},
	\citenamefont {Ouellet-Plamondon}, \citenamefont {\'{E}thier-Majcher},
	\citenamefont {Saint-Jean}, \citenamefont {Andr\'{e}}, \citenamefont {Klem},\
	and\ \citenamefont {Francoeur}}]{Marcet2010}%
\BibitemOpen
\bibfield  {author} {\bibinfo {author} {\bibfnamefont {S.}~\bibnamefont
		{Marcet}}, \bibinfo {author} {\bibfnamefont {C.}~\bibnamefont
		{Ouellet-Plamondon}}, \bibinfo {author} {\bibfnamefont {G.}~\bibnamefont
		{\'{E}thier-Majcher}}, \bibinfo {author} {\bibfnamefont {P.}~\bibnamefont
		{Saint-Jean}}, \bibinfo {author} {\bibfnamefont {R.}~\bibnamefont
		{Andr\'{e}}}, \bibinfo {author} {\bibfnamefont {J. F.}~\bibnamefont {Klem}}, \
	and\ \bibinfo {author} {\bibfnamefont {S.}~\bibnamefont {Francoeur}},\ }\href
{\doibase 10.1103/PhysRevB.82.235311} {\bibfield  {journal} {\bibinfo
		{journal} {Physical Review B}\ }\textbf {\bibinfo {volume} {82}},\ \bibinfo
	{pages} {235311} (\bibinfo {year} {2010})}\BibitemShut {NoStop}%
\bibitem [{\citenamefont {St-Jean}\ \emph {et~al.}(2015)\citenamefont
	{St-Jean}, \citenamefont {Ethier-Majcher},\ and\ \citenamefont
	{Francoeur}}]{St-Jean2015}%
\BibitemOpen
\bibfield  {author} {\bibinfo {author} {\bibfnamefont {P.}~\bibnamefont
		{St-Jean}}, \bibinfo {author} {\bibfnamefont {G.}~\bibnamefont
		{Ethier-Majcher}}, \ and\ \bibinfo {author} {\bibfnamefont {S.}~\bibnamefont
		{Francoeur}},\ }\href {\doibase 10.1103/PhysRevB.91.115201} {\bibfield
	{journal} {\bibinfo  {journal} {Phys. Rev. B}\ }\textbf {\bibinfo {volume}
		{91}},\ \bibinfo {pages} {115201} (\bibinfo {year} {2015})}\BibitemShut
{NoStop}%
\bibitem [{\citenamefont {Hopfield}\ \emph {et~al.}(1966)\citenamefont
	{Hopfield}, \citenamefont {Thomas},\ and\ \citenamefont
	{Lynch}}]{Hopfield1966}%
\BibitemOpen
\bibfield  {author} {\bibinfo {author} {\bibfnamefont {J.}~\bibnamefont
		{Hopfield}}, \bibinfo {author} {\bibfnamefont {D.}~\bibnamefont {Thomas}}, \
	and\ \bibinfo {author} {\bibfnamefont {R.}~\bibnamefont {Lynch}},\ }\href
{http://journals.aps.org/prl/abstract/10.1103/PhysRevLett.17.312} {\bibfield
	{journal} {\bibinfo  {journal} {Physical Review Letters}\ }\textbf {\bibinfo
		{volume} {17}},\ \bibinfo {pages} {312} (\bibinfo {year} {1966})}\BibitemShut
{NoStop}%
\bibitem [{\citenamefont {Francoeur}\ and\ \citenamefont
	{Marcet}(2010)}]{Francoeur2010}%
\BibitemOpen
\bibfield  {author} {\bibinfo {author} {\bibfnamefont {S.}~\bibnamefont
		{Francoeur}}\ and\ \bibinfo {author} {\bibfnamefont {S.}~\bibnamefont
		{Marcet}},\ }\href {\doibase 10.1063/1.3457851} {\bibfield  {journal}
	{\bibinfo  {journal} {Journal of Applied Physics}\ }\textbf {\bibinfo
		{volume} {108}},\ \bibinfo {pages} {043710} (\bibinfo {year}
	{2010})}\BibitemShut {NoStop}%
\end{thebibliography}
\end{document}